\documentclass[sn-mathphys,Numbered]{sn-jnl} 
\usepackage{graphicx}%
\usepackage{multirow}%
\usepackage{amsmath,amssymb,amsfonts}%
\usepackage{amsthm}%
\usepackage{mathrsfs}%
\usepackage[title]{appendix}%
\usepackage{xcolor}%
\usepackage{textcomp}%
\usepackage{manyfoot}%
\usepackage{booktabs}%
\usepackage{algorithm}%
\usepackage{algorithmicx}%
\usepackage{algpseudocode}%
\usepackage{listings}%
\usepackage{natbib}

\usepackage{color}
\usepackage{tensor}
\usepackage{braket}

\begin{document}

\title{Note on episodes in the history of modeling measurements in local spacetime regions using QFT}

\author*[1]{Doreen Fraser}\email{dlfraser@uwaterloo.ca}
\author[2]{Maria Papageorgiou}
\affil*[1]{Department of Philosophy, University of Waterloo}\email{mepapage@uwaterloo.ca}
\affil[2]{Department of Applied Mathematics and Institute for Quantum Computing, University of Waterloo and
Division of Theoretical and Mathematical Physics, University of Patras}

\abstract{The formulation of a measurement theory for relativistic quantum field theory (QFT) has recently been an active area of research. In contrast to the asymptotic measurement framework that was enshrined in QED, the new proposals aim to supply a measurement framework for measurements in local spacetime regions. This paper surveys episodes in the history of quantum theory that contemporary researchers have identified as precursors to their own work and discusses how they laid the groundwork for current approaches to local measurement theory for QFT.}

\maketitle

\noindent \textbf{Acknowledgements} Thank you to Charis Anastopoulos, Jos{\'e} de Ram{\'o}n Rivera, Alex Blum and two anonymous referees for useful feedback on drafts of this paper. This work is part of a larger project for which we are appreciative of the help that we have received from many people (see \cite{pittphilsci22322}). MP is grateful to Bernadette Lessel for co-organising a pandemic reading group that motivated part of this work. DF and MP gratefully acknowledge support from a Social Sciences and Humanities Research Council of Canada Insight Grant. MP acknowledges support of the ID$\#$ 62312 grant from the John Templeton Foundation, as part of the \href{https://www.templeton.org/grant/thequantum-information-structure-ofspacetime-qiss-second-phase}{‘The Quantum Information
Structure of Spacetime’ Project (QISS)}.

\medskip

\noindent \textbf{Author Contributions}  Both authors contributed equally to the research, writing, and revision of the paper. Both authors approved the final manuscript.

\section{Introduction}\label{sec1}

In non-relativistic quantum mechanics (NRQM), predicted quantities typically take the form of properties of an instantaneous state at a finite time or a stationary state. When the dust settled on QED, predictions took the form of scattering amplitudes, which involve asymptotic states in the limit of infinitely early or late times when systems are infinitely far apart and therefore assumed free. This has served well for predicting the outcomes of scattering experiments, but more recently attention has turned to the question of how to model local measurements:  experiments that involve measurements of relativistic quantum fields in local regions of spacetime. For example, this is an important issue in relativistic quantum information, which is devoted to the theoretical and experimental treatment of information-theoretic processes with relativistic quantum systems. In the past few years, formal measurement theories have been developed for local measurements on systems represented using QFT, including \cite{fewster2020quantum,Polo_Gomez_2022,Anastopoulos_2023,oeckl2019local}. Another area in which the representation of local measurements of quantum fields has become a pressing issue is quantum gravity. For example, the recent debate about whether gravitationally-induced entanglement implies that gravity must be quantized also raises questions about how local measurements on quantum fields are modeled and interpreted \cite{PhysRevD.106.076018,PhysRevD.105.086001}. In this note we will aim to put these recent developments in historical context by surveying a few episodes in the history of modeling local measurements on QFT systems.

To appreciate the parallel history of attempts to model local measurements using QFT, it is useful to consider how QED came to be formulated using scattering theory in the first place. Blum \cite{blum_state_2017} offers a comprehensive historical account of this development. He characterizes this historical shift from a focus on states in NRQM to scattering theory in Dyson's formulation of QED as a Kuhnian paradigm shift because it constitutes a significant change in the paradigmatic problem of what is to be calculated from the theory \cite[p.46]{blum_state_2017}. Blum offers an illuminating account of how the quantum state ``withers away"\footnote{It withers away rather than being completely abolished because non-scattering problems (e.g., bound state problems) were solved by starting from the S-matrix and reintroducing stationary or instantaneous states \cite[p.77, fn 81]{blum_state_2017}.} by tracing two lines of development of relativistic quantum theory in the 1930's and 1940's that prepared the ground for the conceptual shift, one that originates with Heisenberg's S-matrix theory and the other with Wheeler-Feynman electrodynamics. As Blum explains, this paradigm shift was prompted both by the desideratum of obtaining an explicitly relativistic formulation of quantum theory and by the need for a calculationally tractable theory. There were correlated shifts in the types of experiments that were important, from spectroscopic experiments to cosmic ray experiments in the 1930s to scattering experiments with particle accelerators \cite[pp.49,78]{blum_state_2017}. At the end of his historical study, Blum raises the following question:

\begin{quote}
But the striking fact that (relativistic) QFT is so readily formulated as a theory of scattering, and that the discovery of this reformulation was so important for the development of the theory, even for its acceptance as a consistent physical theory, calls for an explanation. In what sense is QFT really a theory of scattering processes, and in what sense is the prevalence of scattering problems merely a historical contingency and an effect of good ol’ American pragmatism? (p.78)
\end{quote}

\noindent Blum poses this question to philosophers, but we believe that the scattered (!) history of attempts to model local measurements in QFT provides a provisional answer:  it is possible to  model local measurements using QFT, so QFT is not only a theory of scattering processes. This answer is provisional because the development of a complete measurement theory for QFT is the subject of ongoing research. These problems of honouring relativistic principles such as covariance and the prohibition on superluminal signalling as well as gaining calculational tractability that were overcome using S-matrices in QED remain obstacles that need to be overcome in contemporary attempts to model local measurements using QFT.

The goal of this paper is the modest one of identifying some episodes in this history of treating local measurements using QFT. After 1950, these episodes occurred against the backdrop of the dominant approach of asymptotic, scattering-based treatments of measurement. We chose these episodes because they are cited as precursors in contemporary work on local measurement in QFT. A number of different approaches to modeling measurement in QFT are being actively pursued (e.g., restrictions on allowed operators in \cite{PhysRevD.105.025003}, the positive formalism \cite{oeckl2019local}, or the histories-based Quantum Temporal Probabilities program \cite{Anastopoulos_2023}), but for illustrative purposes we have focused on two well-developed proposals that are much different in spirit:\footnote{For a review of recent approaches to measurement in QFT and how they respond to Sorkin `impossible measurement' scenarios, see \cite{pittphilsci22322}. For a discussion of some philosophical implications of measurement theory for QFT see \cite{Fraser2023}} a detector-based approach pursued in the Relativistic Quantum Information (RQI) community (for example, by Polo-G{\'{o}}mez, Garay and Mart{\'i}n-Mart{\'i}nez in \cite{Polo_Gomez_2022}, and see also \cite{SmithAlexander2017}) and a measurement framework for Algebraic Quantum Field Theory developed in the mathematical physics community by Fewster and Verch in \cite{fewster2020quantum}. In Sec. \ref{ContContext} we give a brief overview of the contemporary work on measurement that provides the starting point for our historical study. Sec. 3 discusses early work by Bohr and Rosenfeld cited in Sorkin \cite{sorkin1993impossible}, which is a seminal thirty-year-old paper that roughly marks the beginning of the recent phase of interest in the problem of representing local measurements in QFT. Fewster and Verch trace the roots of their local measurement theory for Algebraic QFT (AQFT)\cite{fewster2020quantum} back to Haag and Kastler's original axiomatic treatment of AQFT and a series of papers by Hellwig and Kraus in 1969 and 1970, which are reviewed in Sec. \ref{algtradition} and \ref{HellwigKraus}. The Relativistic Quantum Information community has mainly focused on constructing concrete models for particular types of probes interacting with particular types of relativistic quantum systems (e.g., \cite{Hu_2012,PhysRevA.86.012111,PhysRevD.93.024019, PhysRevLett.107.150402}).\footnote{Note that not all of this research is confined to measurements in local regions. For example, sometimes Gaussian smearing functions are used \cite{PhysRevD.108.045015}.} Sec. \ref{UDW model} outlines the origins of this line of research in earlier research on models of detectors, especially the Unruh-DeWitt detector and even earlier models for quantum optics (e.g., \cite{PhysRev.130.2529}). We will conclude with some remarks on how the contemporary work on local measurement in QFT builds on these historical sources. More specifically, we identify some of the research since the 1940s that has made recent progress on modeling local measurements in QFT possible and note that the use of a local version of scattering theory in recent proposals contrasts with both the instantaneous states used in NRQM and the asymptotic scattering theory traditionally used in QED. This short note is intended to provide a starting point for a more systematic historical treatment of attempts to model measurements in local spacetime regions using QFT. A more comprehensive review of contemporary research on local measurement theory for QFT can be found in the companion paper \cite{pittphilsci22322}.

\section{Contemporary Context}\label{ContContext}

Thirty years ago, Sorkin's paper ``Impossible measurements on quantum fields" \cite{sorkin1993impossible} drew attention to the problems that arise when strategies from non-relativistic quantum mechanics are used to model local measurements in QFT. This paper presents examples of `impossible measurements' in which superluminal signalling is permitted when the standard measurement theory for quantum mechanics is applied to describe tripartite measurement scenarios on relativistic systems.

\begin{figure}
\centering
\includegraphics[width=0.8\textwidth]{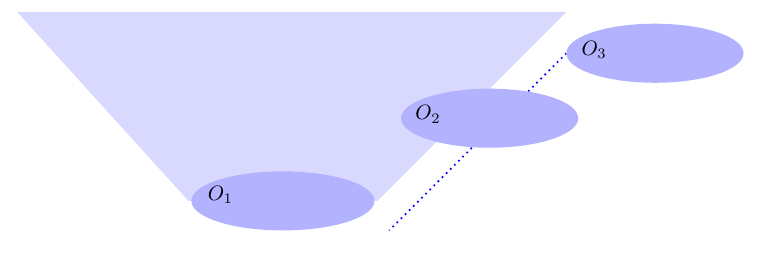}
\hspace{-2cm}
\caption{Region $O_2$ partially invading the future lightcone of $O_1$ and the past lightcone of $O_3$. (Figure based on treatment of `impossible measurements' by Borsten, Jubb, and Kells in \cite{borsten2019impossible}.)}
\label{fig2}
\end{figure}

Sorkin assumes only a minimal, informal framework for relativistic quantum theory. Assume that (when no measurements occur) there is a Heisenberg picture representation of some quantum field $\Phi$ (e.g., a free scalar quantum field) and an observable $A_k$ is associated with a region of Minkowski spacetime $O_k$ by restriction of the field $\Phi$ to $O_k$. Microcausality (i.e., that field observables commute in spacelike separation) is the only principle from QFT that is assumed. This starting point is intended to invoke only generally agreed upon features of relativistic quantum theory. That standard measurement theory (including L{\"u}ders' rule for state update) entails superluminal signalling is demonstrated as follows: a local unitary `kick' $U_{\lambda}= \text{exp}(i\lambda \Phi_1)$ is implemented in a bounded spacetime region $O_1$. Independent of the interpretation of this `local kick', prohibition of superluminal signalling entails that expectation values of observables in a region $O_3$ outside the causal future of $O_1$ should not depend on the value of $\lambda$. In this case of two spacetime regions, this is guaranteed by Microcausality. Nevertheless, Sorkin demonstrates that if a \textit{non-selective} measurement is performed over an `intermediate' spacetime region $O_2$ (which intersects the causal future of $O_1$ and the causal past of $O_3$, see figure \ref{fig2}) then expectation values of observables over region $O_3$ can depend on the magnitude $\lambda$ of the local `kick' over region $O_1$.  Borsten, Jubb, and Kells \cite{borsten2019impossible} recently made important contributions that refine this argument. Such arguments pose the question: which operations are locally implementable in a general QFT scenario without entailing superluminal signalling?

The detector model approach offers a pragmatic answer to this question by considering local operations that are induced by modeling `realistic' detector-field interactions (e.g. in light-matter interaction and quantum optics). Standard quantum measurement theory is applied to `read-out' the detector (modeled as a quantum-mechanical system) and the induced field operations are defined by tracing out the detector degrees of freedom \cite{Polo_Gomez_2022}. In this approach, there is superluminal signalling due to the introduction of non-relativistic detector models, but it is negligible. Specifically, the causality violations can be quantified on a case-by-case basis based on the relevant scales of each set-up (coupling strength, effective size of the interaction etc.) and can be pushed outside the regime of validity of each model \cite{Benincasa_2014,PhysRevD.103.085002}.


The Fewster-Verch (FV) framework for measurement in AQFT \cite{fewster2020quantum,fewster_measurement_2023} takes a different approach to modeling local measurements and to addressing Sorkin's `impossible measurements' problem. The starting point is a choice of axioms for AQFT that is inspired by the modern locally covariant approach to QFT and applies generally to globally hyperbolic spacetimes. This set of axioms includes Microcausality as well as additional axioms that are intended to capture assumptions about relativistic causality. The system, probe,\footnote{A microscopic quantum mechanical system (such as a particle with spin or an atom) is commonly called a \textit{probe} of the field, while the term \textit{detector} is typically used for explicitly macroscopic detector systems (such as a superconducting qubit). Sometimes the terms `detector' and `probe' are used interchangeably, especially if it is not clear from the context whether the measuring system is microscopic or macroscopic.} and their interaction are described using AQFT.  It has been shown that `impossible measurement' scenarios with superluminal signalling are ruled out entirely by the FV framework \cite{PhysRevD.103.025017}.

From a sociological point of view, the detector models approach and FV framework are continuations of different research programs in physics. As we will explain below, the detector models program grew out of attempts to get a handle on the phenomenology of relativistic spacetime physics using models such as the Unruh-DeWitt model. The FV framework is a continuation of the mathematical physics program of algebraic QFT. In recent years, the Relativistic Quantum Information (RQI) series of conferences have been an important venue for discussions about different ways of modeling local measurement in QFT. This includes different predictions that result from modeling measurement in different ways \cite{ruep_weakly_2021,grimmer_measurements_2021}. (See \cite{pittphilsci22322} for a more comprehensive discussion of the `impossible measurements' problem and different approaches to modeling local measurement in QFT.)

\section{Bohr and Rosenfeld (1933) and (1950)}\label{BohrRosenfeld}

Bohr and Rosenfeld's 1933 \cite{bohr1933frage} and 1950 \cite{PhysRev.78.794} papers are two of only five references in Sorkin's `impossible measurements' paper \cite{sorkin1993impossible}. Sorkin mentions them as ``one of the few attempts I know of to design concrete models of field measurements" (p.10). We will first consider the historical context of Bohr and Rosenfeld's papers, and then discuss Sorkin's suggested use of Bohr and Rosenfeld's models.

 The debate about quantum field measurability in the $1930$s was centered around the uncertainty principle or, more generally, Bohr's complementarity. First, Heisenberg attempted to extend the uncertainty principle to a relativistic set up to argue that the limitations on quantum field measurements are analogous to the ones in non-relativistic quantum mechanics \cite{heisenberg1930physical}. Landau and Peierls \cite{landau1931erweiterung} argued that the limitations on quantum measurement are more severe in QFT than in quantum mechanics, challenging the physical basis of the theory. Bohr and Rosenfeld \cite{bohr1933frage} responded to their argument, challenging their assumption of electrically charged pointlike particles as test bodies. Instead, they argue that one must consider spatially extended charged test bodies, whose atomistic structure can be ignored and whose charge density can be adjusted. As a result, the physical predictions of the theory would correspond to field averages over extended spacetime regions, and would not rely on the idealisation of `field at a point'. By controlling the macroscopic charge density of the macroscopic test body one can control the effect of local field fluctuations and, envisioning a suitable compensation mechanism, the spacetime averages of field amplitudes over bounded regions can be measured in principle (up to the limitations that follow from the field commutation relations). The emphasis on the macroscopic aspect of the test body that is `measuring' the quantum field is in line with Bohr's views about quantum measurement.  

The debate between Bohr \& Rosenfeld and Landau \& Peierls about quantum field measurement has been characterised as the ``small war of Copenhagen" \cite{hartz2015uses}. This debate framed much of the discussion about the role of complementarity, the correspondence principle, and the relation of a (not yet settled) mathematical formalism for measurable quantities. From $1936$ to $1946$, with the gradual development and establishment of S-matrix theory by Heisenberg, there is a gradual shift from epistemological to more pragmatic arguments in the spirit of the S-matrix program \cite{blum_state_2017}. After the renormalisation of QED at the end of the 1940s, Bohr and Rosenfeld write the second paper on the measurability of QFT \cite{PhysRev.78.794}, where they review the proposal for measuring field averages over an extended region, and they also propose an idealised arrangement for measuring charge-current densities over the boundary. The proposed arrangement involves a distribution of test bodies over the boundary of a region for measuring the flux. They consider ``the effect of the charge-current density appearing as a consequence of actual or virtual electron
pair production by the field action of the displacement of the test bodies during the measuring process" and determine that ``these effects, which are inseparably connected with the measurements, do not in any way limit
the possibilities of testing the theory" \cite{PhysRev.78.794}. In this second paper they do not put as much emphasis on the macroscopic aspect of the test bodies, since for accurate flux measurements the atomistic structure of the test bodies might come into play \cite[p.411]{cohen2012selected}. The issue of microscopic versus macroscopic, as well as quantum versus classical treatment of the measuring apparatus,  continued to be relevant in the debate surrounding quantum measurement. In the 1960s, Rosenfeld and collaborators (Daneri, Loinger, Prosperi, see \cite{daneri1962quantum}) worked out an account of macroscopic quantum apparatus based on thermodynamical arguments (`irreversibility' of measurement records etc.), arguing against some of the philosophical consequences that followed from von Neumann's (and later Wigner's) account of measurements, such as interpretations of the state `collapse' \cite{jacobsen2012leon}.

Sorkin \cite{sorkin1993impossible} suggests that Bohr and Rosenfeld's proposal for measuring smeared-field amplitudes might provide a testing ground for Sorkin's claim that there are ideal measurement scenarios in QFT  in which superluminal signalling is predicted to occur. Sorkin suggests that the set-up in Bohr and Rosenfeld (1933) could be used to model ideal measurement in his proposed scenario if the classical treatment of the apparatus were replaced by a quantum one ``in order to learn how close they come to actually fulfilling the requirements for an ideal measurement" \cite{sorkin1993impossible}. He elaborates that ``[s]pecifically, one can ask whether they actually measure the field averages they claim to, and whether the probabilities of the different possible outcomes are those predicted by the quantum formalism (with special reference to the use of the projection postulate after the first measurement, since its effect could only be seen in a full quantum treatment)." Sorkin conjectures that the quantum version of Bohr and Rosenfeld's model for ideal measurements on fields in local regions will exhibit superluminal signalling when applied to Sorkin-type scenarios.

Essentially, Sorkin is calling for Bohr and Rosenfeld's models to be modified to fit into the framework for modeling measurements that was introduced by von Neumann and which is now commonly used. It should be noted that Bohr and Rosenfeld's view of quantum measurement is very different in spirit from von Neumann's, in which another microscopic system is coupled to the quantum system that is to be measured. Moreover, Bohr and Rosenfeld intentionally avoided adopting von Neumann's framework for measurement. von Neumann's axiomatisation of quantum theory \cite{vonNeumann1932} laid the groundwork for what is today known as quantum measurement theory. Both works were published in the early 1930s. Rosenfeld matured as a mathematical physicist during his stay at G\"ottingen in 1928-1929 where he got to study what he later called the `Neumanistics' (\cite{jacobsen2012leon}, p.30). He later clarified that Bohr's goal had never been to provide a measurement theory in von Neumann's style \cite{jacobsen2012leon}. It is beyond the scope of this note to elaborate on the differences between Bohr and Rosenfeld's and von Neumann's approaches, but it should be recognized that, contrary to what has been commonly believed, Bohr and Rosenfeld's papers have been influential, albeit in ways that often do not match the original spirit of the work \cite{hartz2015uses}.

\section{Algebraic tradition of operationalism in 1950s and 1960s}\label{algtradition}

One might think that algebraic QFT would have been a context in which the representation of local measurements in QFT was worked out long ago because Haag and others adopted an operational interpretation of the local algebras; however, this was not the case. In a recent paper, Fewster and Verch \cite[p.2]{fewster2020quantum} remark that there has been a ``gap" between the fields of quantum measurement theory and algebraic QFT that ``has--surprisingly--lain open for a long time." They recognize Haag and Kastler's \cite{haagkastler1964} set of algebraic axioms and operationalization as a precursor to their own treatment of local measurement in AQFT. Fewster and Verch adopt a different set of algebraic axioms and generalize their framework to globally hyperbolic spacetimes, but for our purposes what is relevant is their perspective on Haag and Kastler's treatment of measurement. They point out two respects in which it falls short of their goals: Haag and Kastler do not ``set out how exactly one measures an observable or performs an operation within a region of spacetime" and they ``were reluctant to interpret elements of the local algebras as observables (which they considered to arise as limits of local algebra elements)" (p.8). Both of these observations are accurate and have their roots in the historical development of AQFT by Haag.

The first point pertains to the kind of operationalism adopted by Haag and Kastler \cite{haagkastler1964}, which was the first of Haag's papers to use abstract C*-algebras. AQFT maps each open-bounded region $O$ of a spacetime $M$ to an algebra of observables $\mathcal{A}(O)$. An algebraic state $\omega$ is a positive, normalized, linear functional from $\mathcal{A}(O)$ to $\mathbb{C}$. An \textit{operation} is defined mathematically as a non-norm-increasing linear functional that maps the set of states on the algebra $\mathcal{A}$ to itself. This mathematical definition of an operation is used to give local algebras $\mathcal{A}(O)$ an operational interpretation in terms of local laboratory procedures:

\begin{quote}
We must turn now to the physical interpretation, i.e., to the following question: Suppose a specific operation (or state) is defined in terms of a laboratory procedure. How do we find the corresponding element in the mathematical description? For the ``operations" the question is partially answered by the assertion: An operation in the space-time region [$O$] corresponds to an element from [$\mathcal{A}(O)$]. \cite[pp.850--851]{haagkastler1964}
\end{quote} 

\noindent However, this operational interpretation is an abstract schema; particular local operators are not associated with particular concrete measurement operations carried out in a local laboratory. 

A second, related point is that AQFT is not connected to predictions via the direct interpretation of local algebras in terms of laboratory procedures. Instead, the connection to predictions is still made through asymptotic scattering theory. This is a longstanding theme in Haag's approach to QFT that can be traced all the way back to his earliest work in the 1950s (see \cite{Haag2010} for Haag's recollections of this period). The result that came to be known as Haag's theorem is first presented in \cite{Haag1955}. Haag's theorem demonstrated that interactions cannot be consistently represented using the Fock space for a free system. This raised the problem of how to formulate collision theory in a mathematically consistent way. Haag-Ruelle scattering theory \cite{Haag1958,Ruelle1962} was regarded by Haag as the most satisfactory answer to this problem because it could accommodate composite particles. The Haag and Kastler paper \cite{haagkastler1964} opens with a paragraph that directly ties ``observables that can be measured in a spacetime region" to the calculation of ``quantities of direct physical interest such as masses of particles and collision cross sections" by appeal to Haag-Ruelle scattering theory. (The fine print, spelled out in footnote 1:  ``At the present stage this claim is an overstatement, but it is a reasonable extrapolation of results described.") This strategy of interpreting local algebras by relating them to collision cross-sections using scattering theory is a persistent feature of Haag's understanding of AQFT which is still present in his much later book \cite[Sec. III.1, Ch. VI]{haag2012local}.\footnote{Haag and Kastler \cite[p.851]{haagkastler1964} remark that ``[i]n any case it is rather evident that one can construct
a good mathematical representative of a Geiger counter coincidence arrangement using the subalgebras for finite regions." Chapter VI of Haag's book \cite{haag2012local} provides a detailed justification for and qualification of this remark in terms of asymptotic particle representations, focusing on the difficulties raised by superselection rules, the infrared problem, and long-range correlations.}

\section{Hellwig and Kraus circa 1970}\label{HellwigKraus}

Fewster and Verch \cite[p.2]{fewster2020quantum} report that ``[t]he main work on measurement of local observables of which we are aware is due to Hellwig and Kraus." Hellwig and Kraus published a set of three articles \cite{PhysRevD.1.566,hellwigkraus1969,hellwigkraus1970b} in 1969 and 1970 in which they adapted early work on quantum measurement theory to QFT. Two of the papers use early ideas from quantum measurement theory to fill out Haag and Kastler's account of operations in AQFT.

Hellwig and Kraus \cite{PhysRevD.1.566} propose a modification of L{\"u}ders' rule for state update for ideal measurements that is manifestly Lorentz covariant and satisfies the Microcausality assumption (that field observables associated with spacelike separated regions commute). This paper responds to problems with L{\"u}ders' rule in relativistic quantum theory raised by Bloch in \cite{PhysRev.156.1377}. The proposal is that, in the relativistic context, the state update prescribed by L{\"u}ders' rule for measurement of a field property $P(C)$ (where $P$ is a projection operator) only applies in the future and side cones of spacetime region $C$. (Hellwig and Kraus argue that this proposal is physically equivalent to a more elaborate, apparently observer-dependent proposal by Schlieder in \cite{schlieder}.) 

The other two papers \cite{hellwigkraus1969,hellwigkraus1970b} use Ludwig's early work on quantum measurement theory \cite{ludwig} to relate Haag and Kastler's formal concept of an operation to a model of the measurement apparatus. The Haag and Kastler \cite{haagkastler1964} axiomatization of QFT is interpreted as a Heisenberg picture representation of QFT in which (in the absence of external interventions) the state of the system is time-independent and the dynamical evolution of the system is represented by the time-dependence of the algebras of observables. An operation---a change of state that is brought about by an (external) physical apparatus acting on the system---is modeled by Hellwig and Kraus using a unitary, finite-time S-matrix interaction $\mathbf{S}$ between the apparatus and system. Following Ludwig \cite{ludwig}, the combined system-and-apparatus are represented using the product Hilbert space $\mathfrak{h} \otimes \mathfrak{h}^\prime$, it is assumed that the initial states of the system and the apparatus are density matrices, and the state update after measuring some property of the apparatus represented by projection operator $Q^\prime$ is

\begin{equation}
    \mathbf{W} = \frac{\hat{\mathbf{W}}}{Tr \hat{\mathbf{W}}}
\end{equation}

\noindent with $\hat{\mathbf{W}}=(1 \otimes Q^{\prime})\mathbf{S}(W \otimes W^{\prime})\mathbf{S}^{*}(1 \otimes Q^{\prime})$. The first article \cite{hellwigkraus1969} establishes that pure operations (i.e., transformations from pure states to pure states) can be represented mathematically by operators with norm less than one. The second article \cite{hellwigkraus1969} addresses the general case. At the end of \cite{hellwigkraus1969}, Hellwig and Kraus note that their result concerns mathematically possible operations, and that a physical approach would also consider physical restrictions on the initial state of the apparatus, measurable properties of the apparatus, and the interactions between system and apparatus. They then argue that Haag and Kastler \cite{haagkastler1964} can be used to formulate within AQFT the physical requirement that an operation be local, which amounts to imposing the requirement that the interaction between the system and apparatus is restricted to a local spacetime region. 

\section{The use of the Unruh-DeWitt detector model in RQI}\label{UDW model}

In the 1970s, the Unruh \cite{Unruh1977}\footnote{Earman \cite{EARMAN201181} points out that \cite{Unruh1977} was presented at the first Marcel Grossmann Meeting on General Relativity in 1975, and takes this to be the first public presentation of the Unruh effect.} and Hawking \cite{PhysRevD.15.2738} effects focused attention on the thermal states of quantum fields and the exploration of the more exotic settings of curved spacetime and black holes. Particle detectors were introduced to extract particle phenomenology from QFT models. The Unruh-DeWitt detector was introduced in \cite{PhysRevD.14.870,DeWitts}. Unruh \cite{PhysRevD.14.870} modeled a detector as a non-relativistic system consisting of a particle in a box (see also \cite{PhysRevD.29.1047}). DeWitt \cite{DeWitts} introduced the modification of a two-level point monopole system (e.g. a spin) that is coupled to the field over a pointlike trajectory.\footnote{The pointlike model suffers from UV divergences that can be regulated through the introduction of suitable test functions \cite{Schlicht}.}

The Unruh-DeWitt detector model has become a paradigm example in the field of Relativistic Quantum Information (RQI). RQI was born out of the need to merge quantum information theory with relativity theory, and a core commitment of the approach is that relativistic QFT is a necessary ingredient.\footnote{For a first-hand account of the origins of the International Society for Relativistic Quantum Information, see \textit{`How our society came into being'} by Bei-Lok Hu in \url{https://www.isrqi.net}.} See Peres and Terno \cite{RevModPhys.76.93} for a pioneering defense of this approach. RQI describes quantum communication through quantum fields (e.g. \cite{PhysRevLett.114.110505}) and the entanglement structure of QFT by locally coupling multiple detectors to the quantum field (e.g.\cite{Reznik2003,PhysRevD.94.064074}). Many variations upon the Unruh-DeWitt model have been developed and applied to probe many different types of relativistic quantum systems described by QFT (see \cite{pittphilsci22322} and references therein).    In the realm of quantum information, the notion of operations performed in local regions that is informally used in the application of quantum mechanics becomes central. Detector models are introduced to extend the use of QFT from high energy physics to relatively low energy systems probed in quantum information or quantum optics \cite{PhysRev.130.2529,davies1976quantum}. 


\section{Connections to recent developments}\label{Conclusion}

Recent work over the past several years on measurement in QFT has two general goals that are thematic threads running through these episodes. First, the focus is on local, rather than asymptotic, models for measurement. Second, the desideratum is physically meaningful models for measurement that include physically realizable local laboratories, physically reasonable descriptions of the detectors, and only observables that are in principle physically measurable (e.g., exclude observables that could be used to signal superluminally). As Blum \cite{blum_state_2017} documents, a formulation of relativistic quantum theory in terms of instantaneous quantum states could not be obtained in the 1930s and 1940s, prompting a shift to using scattering theory to formulate QED. This is perhaps one reason why Sorkin went all the way back to Bohr and Rosenfeld's 1933 paper \cite{bohr1933frage} to find a potential example of a concrete model of local measurements on quantum fields. Blum also draws attention to the shift in the types of experiments that were regarded as important, from spectroscopic experiments to experiments that were suited to description using scattering theory, such as cosmic ray experiments and scattering experiments at particle accelerators. The recent attention devoted to development of a local measurement theory for QFT has been prompted by another shift in the kinds of experiments that are regarded as important:  quantum information concerns experiments that involve communication using quantum systems on which operations are restricted to local spacetime regions and which are subject to relativistic constraints on allowed processes \cite{PhysRevA.64.052309}.    

The historical episodes sketched in this paper were chosen because they are referenced in recent work on measurement theory for QFT. How does this recent work build on these historical episodes? Sorkin \cite{sorkin1993impossible} presented examples of `impossible measurement' scenarios in which superluminal signalling can occur to draw attention to the fact that standard measurement theory from NRQM cannot be straightforwardly applied to model measurements in QFT. He looked back at Bohr and Rosenfeld's early work in an attempt to find a model of local measurement that might (problematically) predict superluminal signalling. However, Sorkin's `impossible measurement' scenarios are formulated using elements of quantum measurement theory such as L{\"u}ders' rule, which was informed by von Neumann's treatment of measurement rather than Bohr and Rosenfeld's. 

Our first example of a recent proposal for a measurement theory for QFT is the Fewster-Verch (FV) framework \cite{fewster2020quantum}, which models the system, probe, and the interaction between them using algebraic QFT. It is interesting to note that the FV framework retains a scattering map, but following Hellwig and Kraus \cite{hellwigkraus1969,hellwigkraus1970b} it is finite-time rather than asymptotic. Furthermore, in the algebraic framework an algebraic state is not an instantaneous state at a time, but an expectation-valued map that encodes the expectation values of the fields in all local regions. The FV framework generalizes the Hellwig and Kraus model by treating globally hyperbolic spacetimes, relaxing the assumption that the dynamics between system and probe be unitary, and framing the account using the abstract algebraic level of representation rather than concrete Hilbert space representations \cite[p.2]{fewster2020quantum}. Furthermore, the FV framework represents each probe as a physical system, requiring that the probe model satisfy the axioms of AQFT and establishing its localization properties. The developments that occurred in quantum measurement theory from Ludwig \cite{ludwig} to the present are also critical for making the formulation of the FV framework possible. With respect to the other Hellwig and Kraus paper \cite{PhysRevD.1.566}, the FV framework does not incorporate a Lorentz covariant version of L{\"u}ders' rule. More radically, the state update rules in the FV framework do not represent a change of state as occurring at any point or surface in spacetime \cite[p.16]{fewster2020quantum}. (See \cite{pittphilsci22322} for further discussion of these points.)

Our second example of a recently proposed measurement theory is the detector-based measurement theory presented by Polo-G{\'{o}}mez, Garay and Mart{\'i}n-Mart{\'i}nez in \cite{Polo_Gomez_2022}. In this approach, which is situated within RQI, the system is modeled using QFT and it is coupled to a detector that is modeled using NRQM. The set up adopts von Neumann's strategy of modeling measurement by coupling the system measured to another microscopic probe system. The Unruh-DeWitt detector model is taken as a prototype, and variations upon this model as well as different kinds of detector models have been proposed. This approach is motivated by the practical aim of obtaining theoretical representations for realistic experiments on relativistic fields, and by experimental proposals for measuring the Unruh effect \cite{PhysRevLett.125.213603}. 
The detector-based measurement theory in Polo-G{\'{o}}mez et. al. \cite{Polo_Gomez_2022} sets out state update rules for the system. It also retains aspects of the scattering theory framework, but takes finite-time processes and not asymptotic processes as basic. In place of instantaneous states, $n$-point functions that directly involve fields at different times play a central role.\footnote{This is also a feature of histories-based formalisms (see \cite{Anastopoulos_2023, e24010004}).}     

We conclude by emphasizing two themes that emerge from consideration of these historical episodes. First, this recent work on modeling local measurement in QFT builds on research that was not available in the 1940s. Subsequent developments in quantum measurement theory for NRQM played a role in Sorkin's identification of problems encountered modeling local measurements in QFT and supplied a starting point for Fewster and Verch's formulation of measurement theory for AQFT. The exploration of quantum field theories in relativistic settings beyond Minkowski spacetime was a foundation for both the detector models approach (e.g., the paradigm case of the Unruh-DeWitt model) and the FV framework (e.g., axioms for globally hyperbolic spacetimes). The underlying reasons that it has taken many decades to address the problem of how to represent local measurements in QFT is that it requires refined understanding of both the principles of relativistic quantum theory and practical issues surrounding the representation of probes and their interactions with quantum fields in relativistic spacetime. Second, we focused on only two of the current proposals for a local measurement theory for QFT, but it is interesting that neither returns to the instantaneous states that figure in the representation of measurement in NRQM. The FV framework employs algebraic states; in the detector models approach, the $n$-point functions that directly involve fields at different times are the primary means of representing states. Furthermore, neither proposal abandons scattering theory entirely. The goal of modeling measurements that occur in local spacetime regions means that asymptotic scattering no longer plays a role, but a local version of scattering theory is used in both. (See \cite{pittphilsci22322} for further analysis.) Further, more comprehensive study of the historical background of recent proposals for representing local measurement in QFT would be of interest for both the history of physics and the continued development of these research programs in physics.

\bibliography{refs-2.bib}


\begin{thebibliography}{61}
\ifx \bisbn   \undefined \def \bisbn  #1{ISBN #1}\fi
\ifx \binits  \undefined \def \binits#1{#1}\fi
\ifx \bauthor  \undefined \def \bauthor#1{#1}\fi
\ifx \batitle  \undefined \def \batitle#1{#1}\fi
\ifx \bjtitle  \undefined \def \bjtitle#1{#1}\fi
\ifx \bvolume  \undefined \def \bvolume#1{\textbf{#1}}\fi
\ifx \byear  \undefined \def \byear#1{#1}\fi
\ifx \bissue  \undefined \def \bissue#1{#1}\fi
\ifx \bfpage  \undefined \def \bfpage#1{#1}\fi
\ifx \blpage  \undefined \def \blpage #1{#1}\fi
\ifx \burl  \undefined \def \burl#1{\textsf{#1}}\fi
\ifx \doiurl  \undefined \def \doiurl#1{\url{https://doi.org/#1}}\fi
\ifx \betal  \undefined \def \betal{\textit{et al.}}\fi
\ifx \binstitute  \undefined \def \binstitute#1{#1}\fi
\ifx \binstitutionaled  \undefined \def \binstitutionaled#1{#1}\fi
\ifx \bctitle  \undefined \def \bctitle#1{#1}\fi
\ifx \beditor  \undefined \def \beditor#1{#1}\fi
\ifx \bpublisher  \undefined \def \bpublisher#1{#1}\fi
\ifx \bbtitle  \undefined \def \bbtitle#1{#1}\fi
\ifx \bedition  \undefined \def \bedition#1{#1}\fi
\ifx \bseriesno  \undefined \def \bseriesno#1{#1}\fi
\ifx \blocation  \undefined \def \blocation#1{#1}\fi
\ifx \bsertitle  \undefined \def \bsertitle#1{#1}\fi
\ifx \bsnm \undefined \def \bsnm#1{#1}\fi
\ifx \bsuffix \undefined \def \bsuffix#1{#1}\fi
\ifx \bparticle \undefined \def \bparticle#1{#1}\fi
\ifx \barticle \undefined \def \barticle#1{#1}\fi
\bibcommenthead
\ifx \bconfdate \undefined \def \bconfdate #1{#1}\fi
\ifx \botherref \undefined \def \botherref #1{#1}\fi
\ifx \url \undefined \def \url#1{\textsf{#1}}\fi
\ifx \bchapter \undefined \def \bchapter#1{#1}\fi
\ifx \bbook \undefined \def \bbook#1{#1}\fi
\ifx \bcomment \undefined \def \bcomment#1{#1}\fi
\ifx \oauthor \undefined \def \oauthor#1{#1}\fi
\ifx \citeauthoryear \undefined \def \citeauthoryear#1{#1}\fi
\ifx \endbibitem  \undefined \def \endbibitem {}\fi
\ifx \bconflocation  \undefined \def \bconflocation#1{#1}\fi
\ifx \arxivurl  \undefined \def \arxivurl#1{\textsf{#1}}\fi
\csname PreBibitemsHook\endcsname

\bibitem[\protect\citeauthoryear{Papageorgiou and Fraser}{2023}]{pittphilsci22322}
\begin{botherref}
\oauthor{\bsnm{Papageorgiou}, \binits{M.}},
\oauthor{\bsnm{Fraser}, \binits{D.}}:
Eliminating the `impossible': Recent progress on local measurement theory for quantum field theory
(2023).
\url{http://philsci-archive.pitt.edu/22322/}
\end{botherref}
\endbibitem

\bibitem[\protect\citeauthoryear{Fewster and Verch}{2020}]{fewster2020quantum}
\begin{barticle}
\bauthor{\bsnm{Fewster}, \binits{C.J.}},
\bauthor{\bsnm{Verch}, \binits{R.}}:
\batitle{Quantum fields and local measurements}.
\bjtitle{Commun. Math. Phys.}
\bvolume{378}(\bissue{2}),
\bfpage{851}--\blpage{889}
(\byear{2020})
\end{barticle}
\endbibitem

\bibitem[\protect\citeauthoryear{Polo-G{\'{o}}mez et~al.}{2022}]{Polo_Gomez_2022}
\begin{botherref}
\oauthor{\bsnm{Polo-G{\'{o}}mez}, \binits{J.}},
\oauthor{\bsnm{Garay}, \binits{L.J.}},
\oauthor{\bsnm{Mart{\'i}n-Mart{\'i}nez}, \binits{E.}}:
A detector-based measurement theory for quantum field theory.
Physical Review D
\textbf{105}(6)
(2022)
\doiurl{10.1103/physrevd.105.065003}
\end{botherref}
\endbibitem

\bibitem[\protect\citeauthoryear{Anastopoulos et~al.}{2023}]{Anastopoulos_2023}
\begin{barticle}
\bauthor{\bsnm{Anastopoulos}, \binits{C.}},
\bauthor{\bsnm{Hu}, \binits{B.-L.}},
\bauthor{\bsnm{Savvidou}, \binits{K.}}:
\batitle{Quantum field theory based quantum information: Measurements and correlations}.
\bjtitle{Annals of Physics}
\bvolume{450},
\bfpage{169239}
(\byear{2023})
\doiurl{10.1016/j.aop.2023.169239}
\end{barticle}
\endbibitem

\bibitem[\protect\citeauthoryear{Oeckl}{2019}]{oeckl2019local}
\begin{barticle}
\bauthor{\bsnm{Oeckl}, \binits{R.}}:
\batitle{A local and operational framework for the foundations of physics}.
\bjtitle{Advances in Theoretical and Mathematical Physics}
\bvolume{23}(\bissue{2}),
\bfpage{437}--\blpage{592}
(\byear{2019})
\end{barticle}
\endbibitem

\bibitem[\protect\citeauthoryear{Hidaka et~al.}{2022}]{PhysRevD.106.076018}
\begin{barticle}
\bauthor{\bsnm{Hidaka}, \binits{Y.}},
\bauthor{\bsnm{Iso}, \binits{S.}},
\bauthor{\bsnm{Shimada}, \binits{K.}}:
\batitle{Complementarity and causal propagation of decoherence by measurement in relativistic quantum field theories}.
\bjtitle{Phys. Rev. D}
\bvolume{106},
\bfpage{076018}
(\byear{2022})
\doiurl{10.1103/PhysRevD.106.076018}
\end{barticle}
\endbibitem

\bibitem[\protect\citeauthoryear{Danielson et~al.}{2022}]{PhysRevD.105.086001}
\begin{barticle}
\bauthor{\bsnm{Danielson}, \binits{D.L.}},
\bauthor{\bsnm{Satishchandran}, \binits{G.}},
\bauthor{\bsnm{Wald}, \binits{R.M.}}:
\batitle{Gravitationally mediated entanglement: Newtonian field versus gravitons}.
\bjtitle{Phys. Rev. D}
\bvolume{105},
\bfpage{086001}
(\byear{2022})
\doiurl{10.1103/PhysRevD.105.086001}
\end{barticle}
\endbibitem

\bibitem[\protect\citeauthoryear{Blum}{2017}]{blum_state_2017}
\begin{botherref}
\oauthor{\bsnm{Blum}, \binits{A.S.}}:
The state is not abolished, it withers away: {How} quantum field theory became a theory of scattering.
Studies in History and Philosophy of Science Part B: Studies in History and Philosophy of Modern Physics
\textbf{60}
(2017)
\doiurl{10.1016/j.shpsb.2017.01.004}
\end{botherref}
\endbibitem

\bibitem[\protect\citeauthoryear{Jubb}{2022}]{PhysRevD.105.025003}
\begin{barticle}
\bauthor{\bsnm{Jubb}, \binits{I.}}:
\batitle{Causal state updates in real scalar quantum field theory}.
\bjtitle{Phys. Rev. D}
\bvolume{105},
\bfpage{025003}
(\byear{2022})
\doiurl{10.1103/PhysRevD.105.025003}
\end{barticle}
\endbibitem

\bibitem[\protect\citeauthoryear{Fraser}{2023}]{Fraser2023}
\begin{botherref}
\oauthor{\bsnm{Fraser}, \binits{D.}}:
Some philosophical implications of measurement in quantum field theory
(2023).
Unpublished manuscript
\end{botherref}
\endbibitem

\bibitem[\protect\citeauthoryear{Smith}{2017}]{SmithAlexander2017}
\begin{botherref}
\oauthor{\bsnm{Smith}, \binits{A.R.H.}}:
Detectors, reference frames, and time.
PhD thesis,
University of Waterloo
(2017).
\url{http://hdl.handle.net/10012/12618}
\end{botherref}
\endbibitem

\bibitem[\protect\citeauthoryear{Sorkin}{1993}]{sorkin1993impossible}
\begin{bchapter}
\bauthor{\bsnm{Sorkin}, \binits{R.D.}}:
\bctitle{Impossible measurements on quantum fields}.
In: \bbtitle{Directions in General Relativity: Proceedings of the 1993 International Symposium, Maryland},
vol. \bseriesno{2},
pp. \bfpage{293}--\blpage{305}
(\byear{1993})
\end{bchapter}
\endbibitem

\bibitem[\protect\citeauthoryear{Hu et~al.}{2012}]{Hu_2012}
\begin{barticle}
\bauthor{\bsnm{Hu}, \binits{B.L.}},
\bauthor{\bsnm{Lin}, \binits{S.-Y.}},
\bauthor{\bsnm{Louko}, \binits{J.}}:
\batitle{Relativistic quantum information in detectors–field interactions}.
\bjtitle{Classical and Quantum Gravity}
\bvolume{29}(\bissue{22}),
\bfpage{224005}
(\byear{2012})
\doiurl{10.1088/0264-9381/29/22/224005}
\end{barticle}
\endbibitem

\bibitem[\protect\citeauthoryear{Anastopoulos and Savvidou}{2012}]{PhysRevA.86.012111}
\begin{barticle}
\bauthor{\bsnm{Anastopoulos}, \binits{C.}},
\bauthor{\bsnm{Savvidou}, \binits{N.}}:
\batitle{Time-of-arrival probabilities for general particle detectors}.
\bjtitle{Phys. Rev. A}
\bvolume{86},
\bfpage{012111}
(\byear{2012})
\doiurl{10.1103/PhysRevA.86.012111}
\end{barticle}
\endbibitem

\bibitem[\protect\citeauthoryear{H\"ummer et~al.}{2016}]{PhysRevD.93.024019}
\begin{barticle}
\bauthor{\bsnm{H\"ummer}, \binits{D.}},
\bauthor{\bsnm{Mart\'{\i}n-Mart\'{\i}nez}, \binits{E.}},
\bauthor{\bsnm{Kempf}, \binits{A.}}:
\batitle{Renormalized {U}nruh-{D}e{W}itt particle detector models for boson and fermion fields}.
\bjtitle{Phys. Rev. D}
\bvolume{93},
\bfpage{024019}
(\byear{2016})
\doiurl{10.1103/PhysRevD.93.024019}
\end{barticle}
\endbibitem

\bibitem[\protect\citeauthoryear{Sab\'{\i}n et~al.}{2011}]{PhysRevLett.107.150402}
\begin{barticle}
\bauthor{\bsnm{Sab\'{\i}n}, \binits{C.}},
\bauthor{\bsnm{Rey}, \binits{M.}},
\bauthor{\bsnm{Garc\'{\i}a-Ripoll}, \binits{J.J.}},
\bauthor{\bsnm{Le\'on}, \binits{J.}}:
\batitle{Fermi problem with artificial atoms in circuit {QED}}.
\bjtitle{Phys. Rev. Lett.}
\bvolume{107},
\bfpage{150402}
(\byear{2011})
\doiurl{10.1103/PhysRevLett.107.150402}
\end{barticle}
\endbibitem

\bibitem[\protect\citeauthoryear{de~Ram\'on et~al.}{2023}]{PhysRevD.108.045015}
\begin{barticle}
\bauthor{\bsnm{Ram\'on}, \binits{J.}},
\bauthor{\bsnm{Papageorgiou}, \binits{M.}},
\bauthor{\bsnm{Mart\'{\i}n-Mart\'{\i}nez}, \binits{E.}}:
\batitle{Causality and signalling in noncompact detector-field interactions}.
\bjtitle{Phys. Rev. D}
\bvolume{108},
\bfpage{045015}
(\byear{2023})
\doiurl{10.1103/PhysRevD.108.045015}
\end{barticle}
\endbibitem

\bibitem[\protect\citeauthoryear{Glauber}{1963}]{PhysRev.130.2529}
\begin{barticle}
\bauthor{\bsnm{Glauber}, \binits{R.J.}}:
\batitle{The quantum theory of optical coherence}.
\bjtitle{Phys. Rev.}
\bvolume{130},
\bfpage{2529}--\blpage{2539}
(\byear{1963})
\doiurl{10.1103/PhysRev.130.2529}
\end{barticle}
\endbibitem

\bibitem[\protect\citeauthoryear{Borsten et~al.}{2021}]{borsten2019impossible}
\begin{botherref}
\oauthor{\bsnm{Borsten}, \binits{L.}},
\oauthor{\bsnm{Jubb}, \binits{I.}},
\oauthor{\bsnm{Kells}, \binits{G.}}:
Impossible measurements revisited.
Physical Review D
\textbf{104}(2)
(2021)
\doiurl{10.1103/PhysRevD.104.025012}
{\href{https://arxiv.org/abs/1912.06141}{{1912.06141}}}
\end{botherref}
\endbibitem

\bibitem[\protect\citeauthoryear{Benincasa et~al.}{2014}]{Benincasa_2014}
\begin{barticle}
\bauthor{\bsnm{Benincasa}, \binits{D.M.T.}},
\bauthor{\bsnm{Borsten}, \binits{L.}},
\bauthor{\bsnm{Buck}, \binits{M.}},
\bauthor{\bsnm{Dowker}, \binits{F.}}:
\batitle{Quantum information processing and relativistic quantum fields}.
\bjtitle{Classical and Quantum Gravity}
\bvolume{31}(\bissue{7}),
\bfpage{075007}
(\byear{2014})
\doiurl{10.1088/0264-9381/31/7/075007}
\end{barticle}
\endbibitem

\bibitem[\protect\citeauthoryear{de~Ram\'on et~al.}{2021}]{PhysRevD.103.085002}
\begin{barticle}
\bauthor{\bsnm{Ram\'on}, \binits{J.}},
\bauthor{\bsnm{Papageorgiou}, \binits{M.}},
\bauthor{\bsnm{Mart\'{\i}n-Mart\'{\i}nez}, \binits{E.}}:
\batitle{Relativistic causality in particle detector models: Faster-than-light signaling and impossible measurements}.
\bjtitle{Phys. Rev. D}
\bvolume{103},
\bfpage{085002}
(\byear{2021})
\doiurl{10.1103/PhysRevD.103.085002}
\end{barticle}
\endbibitem

\bibitem[\protect\citeauthoryear{Fewster and Verch}{2023}]{fewster_measurement_2023}
\begin{botherref}
\oauthor{\bsnm{Fewster}, \binits{C.J.}},
\oauthor{\bsnm{Verch}, \binits{R.}}:
Measurement in quantum field theory
(2023).
\url{http://arxiv.org/abs/2304.13356}
\end{botherref}
\endbibitem

\bibitem[\protect\citeauthoryear{Bostelmann et~al.}{2021}]{PhysRevD.103.025017}
\begin{barticle}
\bauthor{\bsnm{Bostelmann}, \binits{H.}},
\bauthor{\bsnm{Fewster}, \binits{C.J.}},
\bauthor{\bsnm{Ruep}, \binits{M.H.}}:
\batitle{Impossible measurements require impossible apparatus}.
\bjtitle{Phys. Rev. D}
\bvolume{103},
\bfpage{025017}
(\byear{2021})
\doiurl{10.1103/PhysRevD.103.025017}
\end{barticle}
\endbibitem

\bibitem[\protect\citeauthoryear{Ruep}{2021}]{ruep_weakly_2021}
\begin{barticle}
\bauthor{\bsnm{Ruep}, \binits{M.H.}}:
\batitle{Weakly coupled local particle detectors cannot harvest entanglement}.
\bjtitle{Classical and Quantum Gravity}
(\byear{2021})
\doiurl{10.1088/1361-6382/ac1b08}
{\href{https://arxiv.org/abs/2103.13400}{{2103.13400}}}
\end{barticle}
\endbibitem

\bibitem[\protect\citeauthoryear{Grimmer et~al.}{2021}]{grimmer_measurements_2021}
\begin{barticle}
\bauthor{\bsnm{Grimmer}, \binits{D.}},
\bauthor{\bsnm{Torres}, \binits{B.d.S.L.}},
\bauthor{\bsnm{Mart{\'i}n-Mart{\'i}nez}, \binits{E.}}:
\batitle{Measurements in {QFT}: Weakly coupled local particle detectors and entanglement harvesting}.
\bjtitle{Physical Review D}
\bvolume{104}(\bissue{8}),
\bfpage{085014}
(\byear{2021})
\doiurl{10.1103/PhysRevD.104.085014}
{\href{https://arxiv.org/abs/2108.02794}{{2108.02794}}}
\end{barticle}
\endbibitem

\bibitem[\protect\citeauthoryear{Bohr and Rosenfeld}{1933}]{bohr1933frage}
\begin{barticle}
\bauthor{\bsnm{Bohr}, \binits{N.}},
\bauthor{\bsnm{Rosenfeld}, \binits{L.}}:
\batitle{Zur frage der messbarkeit der elektromagnetshen feldgrossen}.
\bjtitle{Kgl. Danske Vidensk. Selskab. Math.-Fys. Medd}
\bvolume{12},
\bfpage{3}
(\byear{1933})
\end{barticle}
\endbibitem

\bibitem[\protect\citeauthoryear{Bohr and Rosenfeld}{1950}]{PhysRev.78.794}
\begin{barticle}
\bauthor{\bsnm{Bohr}, \binits{N.}},
\bauthor{\bsnm{Rosenfeld}, \binits{L.}}:
\batitle{Field and charge measurements in quantum electrodynamics}.
\bjtitle{Phys. Rev.}
\bvolume{78},
\bfpage{794}--\blpage{798}
(\byear{1950})
\doiurl{10.1103/PhysRev.78.794}
\end{barticle}
\endbibitem

\bibitem[\protect\citeauthoryear{Heisenberg}{1930}]{heisenberg1930physical}
\begin{botherref}
\oauthor{\bsnm{Heisenberg}, \binits{W.}}:
The physical principles of the quantum theory: Transl. into {E}ngl. by {C}arl {E}ckart and {F}rank {C}. {H}oyt
(1930)
\end{botherref}
\endbibitem

\bibitem[\protect\citeauthoryear{Landau and Peierls}{1931}]{landau1931erweiterung}
\begin{barticle}
\bauthor{\bsnm{Landau}, \binits{L.}},
\bauthor{\bsnm{Peierls}, \binits{R.}}:
\batitle{Erweiterung des unbestimmtheitsprinzips f{\"u}r die relativistische quantentheorie}.
\bjtitle{Zeitschrift f{\"u}r Physik}
\bvolume{69}(\bissue{1-2}),
\bfpage{56}--\blpage{69}
(\byear{1931})
\end{barticle}
\endbibitem

\bibitem[\protect\citeauthoryear{Hartz and Freire}{2015}]{hartz2015uses}
\begin{bbook}
\bauthor{\bsnm{Hartz}, \binits{T.}},
\bauthor{\bsnm{Freire}, \binits{O.}}:
\bbtitle{Uses and Appropriations of Niels Bohr’s Ideas About Quantum Field Measurement, 1930--1965}
vol. \bseriesno{1},
pp. \bfpage{397}--\blpage{418}
(\byear{2015})
\end{bbook}
\endbibitem

\bibitem[\protect\citeauthoryear{Cohen and Stachel}{2012}]{cohen2012selected}
\begin{bbook}
\bauthor{\bsnm{Cohen}, \binits{R.S.}},
\bauthor{\bsnm{Stachel}, \binits{J.J.}}:
\bbtitle{Selected Papers of L{\'e}on Rosenfeld}
vol. \bseriesno{21}.
\bpublisher{Springer},
\blocation{\,}
(\byear{2012})
\end{bbook}
\endbibitem

\bibitem[\protect\citeauthoryear{Daneri et~al.}{1962}]{daneri1962quantum}
\begin{barticle}
\bauthor{\bsnm{Daneri}, \binits{A.}},
\bauthor{\bsnm{Loinger}, \binits{A.}},
\bauthor{\bsnm{Prosperi}, \binits{G.M.}}:
\batitle{Quantum theory of measurement and ergodicity conditions}.
\bjtitle{Nuclear physics}
\bvolume{33},
\bfpage{297}--\blpage{319}
(\byear{1962})
\end{barticle}
\endbibitem

\bibitem[\protect\citeauthoryear{Jacobsen}{2012}]{jacobsen2012leon}
\begin{bbook}
\bauthor{\bsnm{Jacobsen}, \binits{A.S.}}:
\bbtitle{Leon Rosenfeld: Physics, Philosophy, and Politics in the Twentieth Century}.
\bpublisher{World Scientific},
\blocation{\hspace{-0.1cm}}
(\byear{2012})
\end{bbook}
\endbibitem

\bibitem[\protect\citeauthoryear{von {N}eumann}{1955}]{vonNeumann1932}
\begin{bbook}
\bauthor{\bsnm{{N}eumann}, \binits{J.}}:
\bbtitle{Mathematical Foundations of Quantum Mechanics}.
\bpublisher{Princeton University Press},
\blocation{\hspace{-0.1cm}}
(\byear{1955}).
\bcomment{First published in German in 1932: Mathematische Grundlagen der Quantenmechank, Berlin: Springer.}
\end{bbook}
\endbibitem

\bibitem[\protect\citeauthoryear{Haag and Kastler}{1964}]{haagkastler1964}
\begin{barticle}
\bauthor{\bsnm{Haag}, \binits{R.}},
\bauthor{\bsnm{Kastler}, \binits{D.}}:
\batitle{An algebraic approach to quantum field theory}.
\bjtitle{J. Mathematical Phys.}
\bvolume{5},
\bfpage{848}--\blpage{861}
(\byear{1964})
\end{barticle}
\endbibitem

\bibitem[\protect\citeauthoryear{Haag}{2010}]{Haag2010}
\begin{barticle}
\bauthor{\bsnm{Haag}, \binits{R.}}:
\batitle{Local algebras. a look back at the early years and at some achievements and missed opportunities}.
\bjtitle{The European Physical Journal H}
\bvolume{35}(\bissue{3}),
\bfpage{255}--\blpage{261}
(\byear{2010})
\end{barticle}
\endbibitem

\bibitem[\protect\citeauthoryear{Haag}{1955}]{Haag1955}
\begin{barticle}
\bauthor{\bsnm{Haag}, \binits{R.}}:
\batitle{On quantum field theories}.
\bjtitle{Danske Vid. Selsk. Mat.-Fys. Medd.}
\bvolume{29}(\bissue{12}),
\bfpage{37}
(\byear{1955})
\end{barticle}
\endbibitem

\bibitem[\protect\citeauthoryear{Haag}{1958}]{Haag1958}
\begin{barticle}
\bauthor{\bsnm{Haag}, \binits{R.}}:
\batitle{Quantum field theories with composite particles and asymptotic conditions}.
\bjtitle{Phys. Rev. (2)}
\bvolume{112},
\bfpage{669}--\blpage{673}
(\byear{1958})
\end{barticle}
\endbibitem

\bibitem[\protect\citeauthoryear{Ruelle}{1962}]{Ruelle1962}
\begin{barticle}
\bauthor{\bsnm{Ruelle}, \binits{D.}}:
\batitle{On the asymptotic condition in quantum field theory}.
\bjtitle{Helv. Phys. Acta}
\bvolume{35},
\bfpage{147}--\blpage{163}
(\byear{1962})
\end{barticle}
\endbibitem

\bibitem[\protect\citeauthoryear{Haag}{2012}]{haag2012local}
\begin{bbook}
\bauthor{\bsnm{Haag}, \binits{R.}}:
\bbtitle{Local Quantum Physics: Fields, Particles, Algebras}.
\bpublisher{Springer},
\blocation{\hspace{-0.1cm}}
(\byear{2012})
\end{bbook}
\endbibitem

\bibitem[\protect\citeauthoryear{Hellwig and Kraus}{1970}]{PhysRevD.1.566}
\begin{barticle}
\bauthor{\bsnm{Hellwig}, \binits{K.-E.}},
\bauthor{\bsnm{Kraus}, \binits{K.}}:
\batitle{Formal description of measurements in local quantum field theory}.
\bjtitle{Phys. Rev. D}
\bvolume{1},
\bfpage{566}--\blpage{571}
(\byear{1970})
\doiurl{10.1103/PhysRevD.1.566}
\end{barticle}
\endbibitem

\bibitem[\protect\citeauthoryear{Hellwig and Kraus}{1969}]{hellwigkraus1969}
\begin{barticle}
\bauthor{\bsnm{Hellwig}, \binits{K.}},
\bauthor{\bsnm{Kraus}, \binits{K.}}:
\batitle{Pure operations and measurements}.
\bjtitle{Communications in Mathematical Physics}
\bvolume{11}(\bissue{3}),
\bfpage{214}--\blpage{220}
(\byear{1969})
\end{barticle}
\endbibitem

\bibitem[\protect\citeauthoryear{Hellwig and Kraus}{1970}]{hellwigkraus1970b}
\begin{barticle}
\bauthor{\bsnm{Hellwig}, \binits{K.}},
\bauthor{\bsnm{Kraus}, \binits{K.}}:
\batitle{Operations and measurements. {II}}.
\bjtitle{Communications in Mathematical Physics}
\bvolume{16}(\bissue{2}),
\bfpage{142}--\blpage{147}
(\byear{1970})
\end{barticle}
\endbibitem

\bibitem[\protect\citeauthoryear{Bloch}{1967}]{PhysRev.156.1377}
\begin{barticle}
\bauthor{\bsnm{Bloch}, \binits{I.}}:
\batitle{Some relativistic oddities in the quantum theory of observation}.
\bjtitle{Phys. Rev.}
\bvolume{156},
\bfpage{1377}--\blpage{1384}
(\byear{1967})
\doiurl{10.1103/PhysRev.156.1377}
\end{barticle}
\endbibitem

\bibitem[\protect\citeauthoryear{Schlieder}{1968}]{schlieder}
\begin{botherref}
\oauthor{\bsnm{Schlieder}, \binits{S.}}:
Some remarks on the change of state of relativistic quantum mechanical systems by measurement and on the locality requirement.
Commun. Math. Phys.
\textbf{7}
(1968)
\end{botherref}
\endbibitem

\bibitem[\protect\citeauthoryear{Ludwig}{1961}]{ludwig}
\begin{bchapter}
\bauthor{\bsnm{Ludwig}, \binits{G.}}:
\bctitle{Gel{\"o}ste und ungel{\"o}ste probleme des me{\ss}prozesses in der quantenmechanik}.
In: \beditor{\bsnm{Bopp}, \binits{F.}} (ed.)
\bbtitle{Werner Heisenberg und die Physik Unserer Zeit},
pp. \bfpage{150}--\blpage{181}.
\bpublisher{Vieweg+Teubner Verlag},
\blocation{Wiesbaden}
(\byear{1961})
\end{bchapter}
\endbibitem

\bibitem[\protect\citeauthoryear{Unruh}{1977}]{Unruh1977}
\begin{bchapter}
\bauthor{\bsnm{Unruh}, \binits{W.G.}}:
\bctitle{Particle detectors and black holes}.
In: \beditor{\bsnm{Ruffini}, \binits{R.}} (ed.)
\bbtitle{Proceedings of the First Marcel Grossmann Meeting on General Relativity},
pp. \bfpage{527}--\blpage{536}.
\bpublisher{North-Holland},
\blocation{Amsterdam}
(\byear{1977})
\end{bchapter}
\endbibitem

\bibitem[\protect\citeauthoryear{Earman}{2011}]{EARMAN201181}
\begin{barticle}
\bauthor{\bsnm{Earman}, \binits{J.}}:
\batitle{The {U}nruh effect for philosophers}.
\bjtitle{Studies in History and Philosophy of Science Part B: Studies in History and Philosophy of Modern Physics}
\bvolume{42}(\bissue{2}),
\bfpage{81}--\blpage{97}
(\byear{2011})
\doiurl{10.1016/j.shpsb.2011.04.001}
\end{barticle}
\endbibitem

\bibitem[\protect\citeauthoryear{Gibbons and Hawking}{1977}]{PhysRevD.15.2738}
\begin{barticle}
\bauthor{\bsnm{Gibbons}, \binits{G.W.}},
\bauthor{\bsnm{Hawking}, \binits{S.W.}}:
\batitle{Cosmological event horizons, thermodynamics, and particle creation}.
\bjtitle{Phys. Rev. D}
\bvolume{15},
\bfpage{2738}--\blpage{2751}
(\byear{1977})
\doiurl{10.1103/PhysRevD.15.2738}
\end{barticle}
\endbibitem

\bibitem[\protect\citeauthoryear{Unruh}{1976}]{PhysRevD.14.870}
\begin{barticle}
\bauthor{\bsnm{Unruh}, \binits{W.G.}}:
\batitle{Notes on black-hole evaporation}.
\bjtitle{Phys. Rev. D}
\bvolume{14},
\bfpage{870}--\blpage{892}
(\byear{1976})
\doiurl{10.1103/PhysRevD.14.870}
\end{barticle}
\endbibitem

\bibitem[\protect\citeauthoryear{DeWitt}{1979}]{DeWitts}
\begin{bchapter}
\bauthor{\bsnm{DeWitt}, \binits{B.}}:
\bctitle{Quantum gravity: the new synthesis}.
In: \beditor{\bsnm{Hawking}, \binits{S.W.}},
\beditor{\bsnm{Israel}, \binits{W.}} (eds.)
\bbtitle{General Relativity: An Einstein Centenary Survey}.
\bpublisher{Cambridge University Press},
\blocation{Cambridge}
(\byear{1979})
\end{bchapter}
\endbibitem

\bibitem[\protect\citeauthoryear{Unruh and Wald}{1984}]{PhysRevD.29.1047}
\begin{barticle}
\bauthor{\bsnm{Unruh}, \binits{W.G.}},
\bauthor{\bsnm{Wald}, \binits{R.M.}}:
\batitle{What happens when an accelerating observer detects a {R}indler particle}.
\bjtitle{Phys. Rev. D}
\bvolume{29},
\bfpage{1047}--\blpage{1056}
(\byear{1984})
\doiurl{10.1103/PhysRevD.29.1047}
\end{barticle}
\endbibitem

\bibitem[\protect\citeauthoryear{Schlicht}{2004}]{Schlicht}
\begin{barticle}
\bauthor{\bsnm{Schlicht}, \binits{S.}}:
\batitle{Considerations on the {U}nruh effect: Causality and regularization}.
\bjtitle{Class. Quant. Grav.}
\bvolume{21},
\bfpage{4647}--\blpage{4660}
(\byear{2004})
\end{barticle}
\endbibitem

\bibitem[\protect\citeauthoryear{Peres and Terno}{2004}]{RevModPhys.76.93}
\begin{barticle}
\bauthor{\bsnm{Peres}, \binits{A.}},
\bauthor{\bsnm{Terno}, \binits{D.R.}}:
\batitle{Quantum information and relativity theory}.
\bjtitle{Rev. Mod. Phys.}
\bvolume{76},
\bfpage{93}--\blpage{123}
(\byear{2004})
\doiurl{10.1103/RevModPhys.76.93}
\end{barticle}
\endbibitem

\bibitem[\protect\citeauthoryear{Jonsson et~al.}{2015}]{PhysRevLett.114.110505}
\begin{barticle}
\bauthor{\bsnm{Jonsson}, \binits{R.H.}},
\bauthor{\bsnm{Mart\'{\i}n-Mart\'{\i}nez}, \binits{E.}},
\bauthor{\bsnm{Kempf}, \binits{A.}}:
\batitle{Information transmission without energy exchange}.
\bjtitle{Phys. Rev. Lett.}
\bvolume{114},
\bfpage{110505}
(\byear{2015})
\doiurl{10.1103/PhysRevLett.114.110505}
\end{barticle}
\endbibitem

\bibitem[\protect\citeauthoryear{Reznik}{2003}]{Reznik2003}
\begin{barticle}
\bauthor{\bsnm{Reznik}, \binits{B.}}:
\batitle{Entanglement from the vacuum}.
\bjtitle{Found. Phys.}
\bvolume{33}(\bissue{1}),
\bfpage{167}--\blpage{176}
(\byear{2003})
\doiurl{10.1023/A:1022875910744}
\end{barticle}
\endbibitem

\bibitem[\protect\citeauthoryear{Pozas-Kerstjens and Mart\'{\i}n-Mart\'{\i}nez}{2016}]{PhysRevD.94.064074}
\begin{barticle}
\bauthor{\bsnm{Pozas-Kerstjens}, \binits{A.}},
\bauthor{\bsnm{Mart\'{\i}n-Mart\'{\i}nez}, \binits{E.}}:
\batitle{Entanglement harvesting from the electromagnetic vacuum with hydrogenlike atoms}.
\bjtitle{Phys. Rev. D}
\bvolume{94},
\bfpage{064074}
(\byear{2016})
\doiurl{10.1103/PhysRevD.94.064074}
\end{barticle}
\endbibitem

\bibitem[\protect\citeauthoryear{Davies}{1976}]{davies1976quantum}
\begin{bbook}
\bauthor{\bsnm{Davies}, \binits{E.B.}}:
\bbtitle{Quantum Theory of Open Systems}.
\bpublisher{Academic Press},
\blocation{\hspace{-0.1cm}}
(\byear{1976})
\end{bbook}
\endbibitem

\bibitem[\protect\citeauthoryear{Beckman et~al.}{2001}]{PhysRevA.64.052309}
\begin{barticle}
\bauthor{\bsnm{Beckman}, \binits{D.}},
\bauthor{\bsnm{Gottesman}, \binits{D.}},
\bauthor{\bsnm{Nielsen}, \binits{M.A.}},
\bauthor{\bsnm{Preskill}, \binits{J.}}:
\batitle{Causal and localizable quantum operations}.
\bjtitle{Phys. Rev. A}
\bvolume{64},
\bfpage{052309}
(\byear{2001})
\doiurl{10.1103/PhysRevA.64.052309}
\end{barticle}
\endbibitem

\bibitem[\protect\citeauthoryear{Gooding et~al.}{2020}]{PhysRevLett.125.213603}
\begin{barticle}
\bauthor{\bsnm{Gooding}, \binits{C.}},
\bauthor{\bsnm{Biermann}, \binits{S.}},
\bauthor{\bsnm{Erne}, \binits{S.}},
\bauthor{\bsnm{Louko}, \binits{J.}},
\bauthor{\bsnm{Unruh}, \binits{W.G.}},
\bauthor{\bsnm{Schmiedmayer}, \binits{J.}},
\bauthor{\bsnm{Weinfurtner}, \binits{S.}}:
\batitle{Interferometric {U}nruh detectors for bose-einstein condensates}.
\bjtitle{Phys. Rev. Lett.}
\bvolume{125},
\bfpage{213603}
(\byear{2020})
\doiurl{10.1103/PhysRevLett.125.213603}
\end{barticle}
\endbibitem

\bibitem[\protect\citeauthoryear{Anastopoulos and Savvidou}{2022}]{e24010004}
\begin{botherref}
\oauthor{\bsnm{Anastopoulos}, \binits{C.}},
\oauthor{\bsnm{Savvidou}, \binits{N.}}:
Quantum information in relativity: The challenge of {QFT} measurements.
Entropy
\textbf{24}(1)
(2022)
\doiurl{10.3390/e24010004}
\end{botherref}
\endbibitem

\end{thebibliography}

\end{document}